# LP-Based Power Grid Enhancement Methodology

Tapio Bohn, *Member, IEEE,* Paul Salmi, *Member, IEEE,* Albert Milner, *Senior Member, IEEE*

**Abstract**
**In this paper, we explored the opportunity to enhance power grid robustness after routing stage, and propose a linear programming based algorithm that maximizes the improvement of power grid strengthening with given available routing resource. We further discussed some techniques to leverage trade-offs between runtime and optimality of the solutions. Experimental results show substantial power integrity improvement with "zero cost".**

## 1. Introduction

In modern digital circuit design, frequency is increasing constantly, power density is increasing drastically, while routing resource does not scale the same way [1]. Thus power grid design has been increasingly challenging since more routing resource need to be saved for signal routing. Power integrity (mainly IR drop and electromigration, i.e. EM) has become extremely difficult to close [2]. In this paper, we propose a post-route power grid enhancement methodology that leverage the existing no-used routing resource to strengthen the power grid to achieve power integrity sign-off quality.

## 2. Motivation

When timing is closed for a given design, there is no need to utilize any of the open routing space anymore [3]. However, these open spaces can be used to strengthen power grid weakness for "zero cost" [4]. Figure 1 illustrates a potential example, how an open space can help improve power integrity of the design.

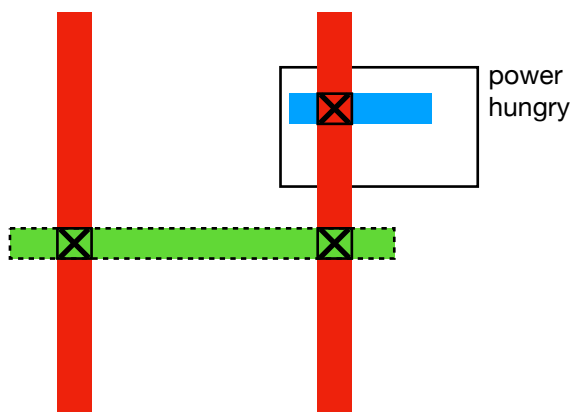

**Fig 1. Using open space to share current**

In Figure 1, an instance is power hungry and is drawing large current that causes IR drop hotspot [5]. By utilizing the open space, we manage to share current from the two adjacent parallel power wires and resolve the hotspot issue [6]. Using this mechanism, it can also help to significantly reduce impact from variations [7][8], especially for critical designs [9]. EM-sensitive wires can also benefits from this operations [10]. In the next Section 3, we will discuss the details of applying a linear programming (LP) based algorithm to optimize the power grid strengthening methodology; in Section 4, we will discuss a few techniques to trade-off between optimality and

## 3. LP-Based Power Grid Enhancement

When using open space, we noticed that there will be potential conflict and competing between two different power wires. Figure 2 illustrates such an example.

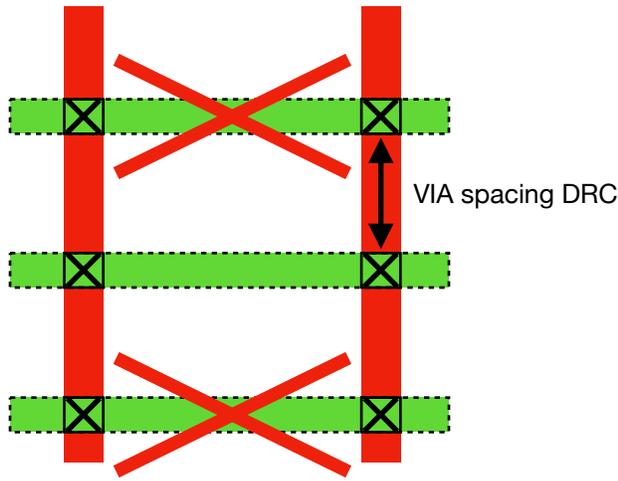

**Fig 2. Potential competing condition**

In Figure 2, when the horizontal center wire is added, the wires above and below will lose its resource, since the via to via spacing design-rule-check (DRC) prevents a via to be created for those wires [11].

Vias are critical to power grid integrity as well as reliability [12]. To optimize the number of wires and vias added to share power supply current, we formulated the problem into a binary linear programming algorithm [13]. The main constraints are coming from via spacing. Equation set (1) defines the problem formulation.

where $i$ is the row index and $j$ is the column index, $x_{i,j}$ is the potential via candidate at location (i,j), $y$ is the total number of vias to be added. There are various numerical solvers to calculate binary linear programming results. However, runtime complexity is a concern, we will discuss technique to speed up the runtime performance in Section 4.

## 4. Trade-Off Technique

In Section 3, we discussed the binary linear programming algorithm to maximize the chances of adding current sharing wires. It should be noted that binary linear programming is a NP-hard problem, and cannot be efficiently solved within polynomial runtime [14]. Even with efficient numerical computation packages in Python or Matlab, runtime is not desirable. To overcome this issue, we propose to split the whole design into separate partitions [15]. Figure 3 illustrates how a break is introduced to split the design and decompose the problem into smaller sub-problems.

Since the runtime exponentially increases with the number of potential via candidates, by artificially split the design into smaller areas, we cut down

$$\text{maximize} \quad y = \sum_{i=1}^{m}\sum_{j=1}^{n} x_{i,j}$$

$$\text{subject to} \quad x_{i,j} + x_{i+1,j} \leq 1, i \in \{1,2,...,m\}, j \in \{1,2,...,n\} \quad (1)$$

$$x_{i,j} \in \{0,1\}$$

the problem sizes. The optimality loss from this technique is small, equation (2) gives the bound of the heuristic solutions using this technique.

$$y' \geq y^* - k \times m \quad (2)$$

where k is the number of break lines introduced, m is the number of rows in the design, y* is the optimal solution, and y' is the suboptimal solution derived using our heuristic technique.

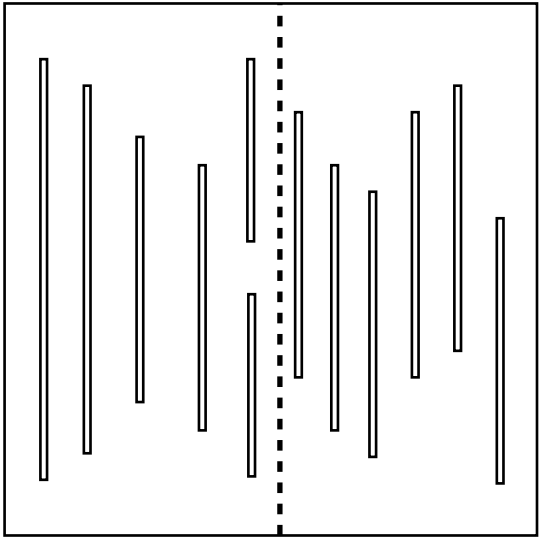

Fig 3. Break lines for partitioning design

## 5. Experiments

We used one block from the open source design or1200_fpu_arith [16] as our benchmark testcases. The logical synthesis is done using Synopsis Design Compiler [17] and physical synthesis is done using Synopsis ICC2 [18]. Sign-off extraction is done using Synopsis Star-RC [19] and power integrity checks are run using Apache RedHawks [20]. Temperature effects are also taken into account [21]. Worst case current is estimated using [22]. Table 1 shows the results of applying our proposed power grid enhancement algorithm.

It can be observed that using our approach of 10 partition LP, we are able to improve the power integrity by 6.0% compared to reference and runtime by 62.6% compared to optimal LP. If we use 100 partition LP, we can achieve 2.1% worst case power integrity improvement and 87.8% runtime improvement compared to optimal LP.

## 6. Conclusions

In this paper we discuss opportunities to enhance power grid after timing closure. By leveraging the open space available, we were able to improve the power integrity significantly with "zero cost". We propose the binary linear programming algorithm to optimize the power grid enhancement result. Given the NP-hard property of binary linear programming algorithm, we proposed a decompose technique to split and reduce problem

Table 1. Comparison between sign-off and DFS checks

|  | reference | optimal LP | 10 partition LP | 100 partiton LP |
|---|---|---|---|---|
| # vias added | N/A | 105.6k | 100.1k | 76.4k |
| voltage drop avg (mV) | 35.6 | 30.8 | 31.2 | 34.1 |
| voltage drop worst (mV) | 66.7 | 62.7 | 63.2 | 65.3 |
| runtime (hrs) | N/A | 12.3 | 4.6 | 1.5 |

size. The split approach greatly reduces the runtime with little loss of optimality.